\documentclass[twocolumn,aps,prl,showpacs]{revtex4}
\usepackage{graphicx}
\usepackage{amsmath}

\begin{document}

\title{Dirac spectra and edge states in honeycomb plasmonic lattices}
\author{Dezhuan Han,$^1$ Yun Lai,$^1$ Jian Zi,$^2$ Zhao-Qing Zhang,$^1$ and
C. T. Chan$^1$}
\affiliation{$^1$Department of Physics, Hong Kong University of Science and Technology,
Clear Water Bay, Kowloon, Hong Kong, China\\
$^2$Department of Physics, Fudan University, Shanghai 200433, People's
Republic of China}
\date{\today}

\begin{abstract}
We study theoretically the dispersion of plasmonic honeycomb
lattices and find Dirac spectra for both dipole and quadrupole
modes. Zigzag edge states derived from Dirac points are found in
ribbons of these honeycomb plasmonic lattices. The zigzag edge
states for out-of-plane dipole modes are closely analogous to the
electronic ones in graphene nanoribbons. The edge states for
in-plane dipole modes and quadrupole modes, however, have rather
unique characters due to the vector nature of the plasmonic
excitations. The conditions for the existence of plasmonic edge
states are derived analytically.
\end{abstract}

\pacs{42.70.Qs, 41.20.Jb, 73.20.Mf}
\maketitle

Metal nanoparticles support remarkable optical resonances known as
localized surface plasmons \cite{kre:95,mai:07}. Surface plasmons
are collective electronic density waves bound at metal surfaces that
can locally enhance incident optical fields by orders of magnitude.
The enhanced optical fields in the near field give rise to many
extraordinary optical phenomena such as surface-enhanced Raman
scattering \cite{nie:97,kne:97} and single-molecule fluorescence
detection \cite{ang:06}. The arrangement of metal nanoparticles
regularly can offer more. It has been shown that a chain of closely
spaced metal nanoparticles can guide electromagnetic (EM) energy
with lateral mode confinement below the diffraction limit
\cite{mai:03}. Two-dimensional (2D) lattices consisting of metal
nanoparticles show interesting tunable optical response over the
entire visible range \cite{tao:07}. A three-dimensional arrangement
of metal nanoparticles can form metallic photonic crystals (PCs)
with robust photonic band gaps that depend on the local order rather
than on the symmetry or the global long range order \cite{zhang:00}.
Metal nanoparticles thus manifest promising building blocks for
plasmonic materials, leading to unprecedented applications in
nanophotonics, nonlinear optics, bio-sensing, and even medical
therapy.

Recently, graphene has received considerable interest because of the
existence of the Dirac point that offers remarkable electronic properties
\cite{nov:04,loui:06,nak:96}. The Dirac spectra in 2D PCs for EM waves have
also been identified \cite{mara:91,hal:08} and the Dirac-point-derived edge
states have novel transportation properties when time-reversal symmetry
breaking is introduced. These edge states can render "one-way" waveguiding
for EM waves, confirmed by numerical simulations \cite{wang:08}. Both
graphene and 2D PCs support Dirac dispersions. However, the Dirac spectrum
in graphene is derived directly from the nearest-neighbor hopping of bound
electron states, while that in 2D PCs corresponds to the photonic bands of
scattering photons.

In this paper, we study theoretically the plasmonic band structures
of metal nanospheres arranged in a honeycomb lattice. The plasmonic
lattice is an open system that supports both guided modes and leaky
modes \cite{park:04}, and as such, it is different from electronic
graphene and 2D PCs. Furthermore, the Dirac dispersions found in
graphene and 2D PCs are only for scalar waves. The 2D plasmonic
lattice can support both scalar waves and vector waves. Our results
reveal that for both scalar and vector waves, this honeycomb
plasmonic lattice possesses Dirac points for the infinite system and
guided edge states when an edge is present. The existence of Dirac
points and edge states in plasmonic systems may open up new avenues
in both physics and applications for the emerging field of
plasmonics.

The honeycomb plasmonic structure under study is shown in Fig. 1(a), in
which each black dot in the $x-y$ plane represents a metal sphere. Its
Brillouin zone (BZ) is shown Fig. 1(b). The inter-distance between adjacent
spheres is $a_{0}=a/\sqrt{3}$ , where $a$ is the lattice constant. The
primitive and reciprocal lattice vectors are, respectively, given by $%
\mathbf{a}_{1,2}=a(\pm 1/2,\sqrt{3}/2)$ and $\mathbf{b}_{1,2}=2\pi (\pm 1,1/%
\sqrt{3})/a$. We use the Drude-type permittivity $\varepsilon
(\omega )=1-\omega _{p}^{2}/\omega ^{2}$ with $\omega _{p}=6.18$ eV
for metal spheres. The lattice constant is chosen to be $a=60$ nm,
and the sphere radius is $r_{s}=10$ nm. For this ratio
$r_{s}/a=\sqrt{3}/6$, the dipole and quadruple dispersions are
separated in energy as demonstrated below.

\begin{figure}[tbp]
\centerline{\includegraphics[width=7.5cm]{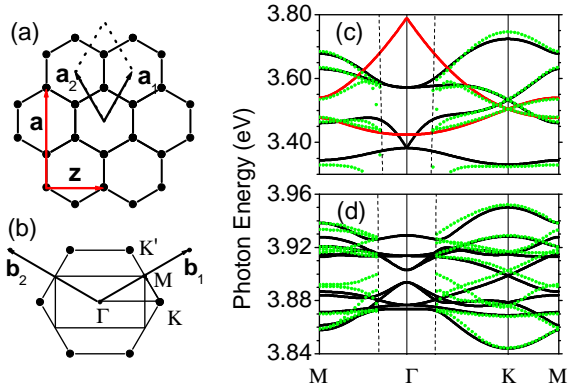}}
\caption{(Color online) (a) Schematic structure of the honeycomb plasmonic
lattice. $\mathbf{a}_{1}$, $\mathbf{a}_{2}$ are the primitive vectors;
\textbf{a} and \textbf{z} are the 1D primitive vectors for armchair and
zigzag ribbons respectively. (b) Reciprocal lattice vectors. (c) QSA band
structures for in-plane dipole modes (black dots) and out-of-plane dipole
modes (red dots); (d) QSA band structures for quadrupoles. MST results
(electrodynamic calculation with retardation effects taken fully into
account) are also shown in (c) and (d) by green dots. The dashed lines in
panels (c) and (d) indicate the light lines. The lattice constant is 60 nm
and the sphere radius is 10 nm.}
\label{fig:fig1}
\end{figure}

The plasmonic dipole dispersions calculated within the quasi-static
approximation (QSA) limit are shown in Fig. 1(c), in which the red dots are
for out-of-plane $\mathbf{P}_{out}$ modes and the black dots correspond to
in-plane $\mathbf{P}_{in}$ modes. The $\mathbf{P}_{in}$ and $\mathbf{P}%
_{out} $ modes form their own orthogonal subspaces. We see that
there are two Dirac cones at the $K$ point, one for the
$\mathbf{P}_{in}$ and one for the $\mathbf{P}_{out}$ modes. Fig.
1(d) shows the plasmonic dispersions in the QSA limit if only
quadrupoles ($\mathbf{Q}$) are considered. We found a Dirac point at
about 3.9 eV, which is the single sphere quadruple resonance
frequency. At these Dirac points, the constant frequency surfaces
are represented by two cones meeting at the $K$ point. It should be
pointed out that the in-plane dipole and quadrupole modes are vector
waves in nature. The existence of Dirac dispersions in these cases
is a result of the symmetry of the honeycomb structure which is
robust against their vector nature.

We note that the $l$-th multipole plasmon resonance frequency of a
small plasmonic sphere is $\omega _{l}=\left[ l/(2l+1)\right]^{1/2}
\omega _{p}$. When the spheres are arranged in a lattice, each
multipole resonance couples to form a set of bands, and these
manifolds do not mix when $r_{s}/a_{0}$ is small. The dipole and
quadrupole bands shown in Fig. 1 are indeed separated in energy. In
Figs. 1(c) and (d), we compare the QSA results with the fully
fledged multiple scattering theory (MST) results (shown in green
color) \cite{fung:07}. The MST gives a solution to Maxwell equations
that takes into account the full retardation effects. The comparison
between the MST (green dots) and QSA results is shown in Fig. 1(c)
for dipole modes and in Fig. 1(d) for quadrupole modes. The details
of the MST method are described elsewhere \cite{fung:07}. The MST
dispersions are shown only for $k>\omega /c$, since the leaky modes
inside the light cone are not well defined due to the coupling with
free photons. As the $K$ point is far away from light cone, the
Dirac dispersion found by QSA agrees well with the MST calculations.
We will discuss the Dirac dispersions with QSA since it can give a
more straightforward interpretation.

An important consequence of the Dirac spectrum in graphene electronics is
the existence of peculiar edge states \cite{nak:96} with unusual physical
properties \cite{loui:06} in graphene ribbons with finite widths. Two types
of graphene nanoribbons, namely armchair and zigzag ribbons, are usually
considered. In Fig. 1(a), the vector $\mathbf{a}$ ($\mathbf{z}$) indicates
the translation vector of the armchair (zigzag) ribbons, and the shorter
(longer) side of the rectangle inside the 2D graphene BZ is the 1D BZ of the
armchair (zigzag) ribbons. The band structure of the armchair (zigzag)
graphene ribbons can be predicted by the projection of 2D graphene bands
onto the corresponding axis of the reciprocal lattice vector. The original
Dirac point $K$ ($K^{\prime }$) is expected to appear at $ka=0$ $(2\pi /3)$
for armchair (zigzag) ribbons. Compared with the projected 2D electronic
graphene bands, a new feature of zigzag ribbons is the appearance of edge
states that extends from $ka=\pi $ to $2\pi /3$, where $ka=2\pi /3$ is the
expected degeneracy point of the 2D projected graphene bands. These new
additional edge states form a flat band at energy $E=0$ with wave functions
localized near the edge sites. Within a tight-binding model, the necessary
condition for $E=0$ edge states is "zero-hopping", namely, the total sum of
hopping terms over the nearest neighbor sites should vanish. For a
semi-infinite graphene sheet with a zigzag edge, the tight-binding model
shows that when $ka=\pi $, the edge state wave function is completely
localized. Thereafter the decay length becomes longer and smoothly connects
to a delocalized state at $ka=2\pi /3$, which is the 2D graphene Dirac point.

\begin{figure}[tbp]
\centerline{\includegraphics[width=3.2in]{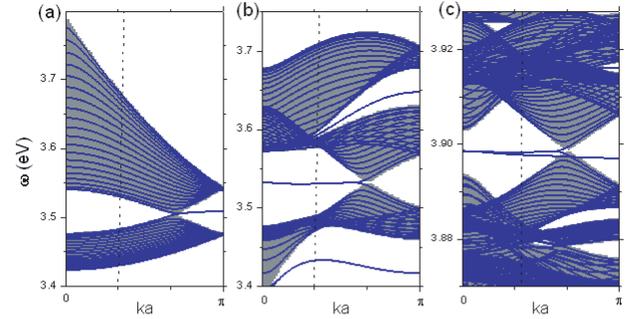}} \caption{(Color
online) Band structures for $N=20$ zigzag plasmonic ribbons (blue)
and projected bands for the 2D honeycomb structure along the
corresponding reciprocal axis (grey area). (a) Out-of-plane dipole
modes. The flat band between $2\protect\pi /3$ and $\protect\pi $
corresponds to edge states. (b) In-plane dipole modes. The flat band
with $ka$ ranging from 0 to $2\protect\pi /3$ corresponds to
Dirac-point-derived edge states. (c) Qudrupole modes. Near the
single particle resonance 3.9 eV ($\protect \omega _{l=2}$), there
are two additional bands: one with $ka$ ranging from 0 to
$2\protect\pi /3$, the other traversing the whole BZ. The system
parameters are the same as in Fig.~\protect\ref{fig:fig1}.}
\label{fig:fig2}
\end{figure}

We now discuss the Dirac-point-derived edge states of plasmonic
ribbons. Unless otherwise stated, the results for finite-width
plasmonic ribbons here are calculated for $N=20$ zigzag and armchair
ribbons, where $N$ is an integer characterizing the width of the
ribbon \cite{nak:96}. We expect that the edge states of out-of-plane
modes might be similar to those of electronic edge states derived
from the out-of-plane $\pi $ states in graphene, while the edge
states for the $\mathbf{P}_{in}$ and $\mathbf{Q}$ modes (if there
are any) might be very different. We note that electronic graphene
\cite{nak:96} and 2D PCs studied recently \cite{hal:08} are all
dealing with scalar waves. In their cases, the electron and photon
transport near the Dirac points can be described by the
"two-component" massless Dirac Hamiltonian $\mathbf{H}=\gamma
(k_{x}\sigma _{x}+k_{y}\sigma _{y})$, where $ \gamma $ is a band
parameter and $\mathbf{\sigma} $ is the Pauli matrix. In the
plasmonic sphere arrays, the vector nature of EM waves enters
explicitly, and the underlying physics can go beyond the
"two-component" Dirac equation.

In Fig. 2, the blue dots represent the plasmonic bands of zigzag
ribbons for out-of-plane dipole, in-plane dipole and quadrupole. The
bands are superimposed on top of the plasmonic bands of the 2D
honeycomb lattice projected along the zigzag axis, which are shown
as the grey background. For $\mathbf{P}_{out}$ modes in Fig. 2(a),
we see that the ribbon bands basically overlap the projected 2D
bands with the exception of an additional degenerate flat band
starting from about $ka=2\pi /3$ to $\pi $ in the gap of the
projected bands, which corresponds to the edge states in the zigzag
graphene ribbons. This flat band occurs at 3.51 eV (the single
sphere dipole resonance $\omega _{l=1}=3.52$ eV), which has the
similar physical meaning of $E=0$ ("zero-hopping") in graphene. The
slight differences in the edge state frequencies and $\omega _{l=1}$
come from the hopping among farther sites in the lattice, which is
neglected in the tight-binding model for graphene.

\begin{figure}[tbp]
\centerline{\includegraphics[width=3.2in]{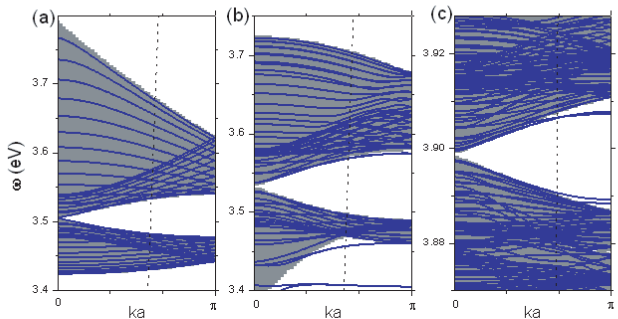}}
\caption{(Color online) Same as Fig. 2 but for armchair plasmonic ribbons.
There is no edge state for the armchair ribbon near the single particle
resonance at 3.52 eV (dipole) or 3.9 eV (quadrupole).}
\label{fig:fig3}
\end{figure}

For $\mathbf{P}_{in}$ modes, there is no edge state from $ka=2\pi/3$
to $\pi$, however, a new additional flat band with its frequency
near $\omega_{l=1}$ emerges from $ka=0$ to about $2\pi/3$. This
phenomenon is rather different from that in the zigzag graphene
ribbon, where the edge states are forbidden near the $\Gamma$ point.
The "two-component" effective Dirac equation commonly used in
graphene electronics is not valid in this case since the vector
nature of EM waves introduces extra degrees of freedom. More
interesting observation can be made when we come to qudrupoles. Two
sets of edge states appear near the single particle resonance at
$\omega_{l=2}=3.9$ eV. The dispersion of one set of edge states is
similar to the in-plane dipole edge mode discussed above, as it is
almost flat and is confined between $0\leq ka < 2 \pi /3$. There is
an additional band of edge states with a small and negative group
velocity that goes all the way from $ka=0$ to $\pi$.

The $\mathbf{P}_{out}$, $\mathbf{P}_{in}$ and $\mathbf{Q}$ modes for
the armchair plasmonic ribbons are shown in Fig. 3 as blue dots,
overlaid on top of the 2D plasmonic bands (grey color) projected
along the armchair axis. There is no additional flat band near the
single sphere resonance $\omega _{l=1}$ for dipoles or $\omega
_{l=2}$ for quadrupoles. We can thus conclude that the
Dirac-point-derived edge states observed in the plasmonic system
come from the boundary condition (zigzag edge) which is analogous to
that in electronic graphene.

In the following, we give an analytic derivation of the condition
for the existence of plasmonic zigzag edge states near multipole
resonance frequencies by truncating the Green function to the
nearest neighbor. Let us consider a general tensor field
$\mathbf{F}$ of rank $l$ (which can be the electric field in the
case of a dipole or the field gradient in the case of a quadrupole)
generated by a localized source $\mathbf{V}$ with rank $l$ (dipole,
quadrupole \ldots ). In free space, the field and the source are
related by the $2l$-th rank Green function $\mathbf{G}(\mathbf{r}-\mathbf{r%
}^{\prime })$ which satisfies $\mathbf{F}(\mathbf{r})=\mathbf{G}(\mathbf{r}-%
\mathbf{r}^{\prime })\mathbf{V}(\mathbf{r}^{\prime })$ and the linear
response gives $\mathbf{V}(\mathbf{r})=\alpha \mathbf{F}(\mathbf{r})$, where
$\alpha $ is the multipole polarizability. We place these point sources $%
\mathbf{V}$ on a semi-infinite honeycomb lattice with a zigzag edge as shown
in the inset of Fig. 4. The indices $(m,n)$ in the inset are used to label
the honeycomb sites, where $n$ and $m$ label the position along (or $y$%
-axis) and perpendicular to ($x$-axis) the zigzag chain respectively
(see Fig. 4). $\mathbf{r}_{m,n}$ is the coordinate of the $(m,n)$
site, and the site $(1,n+1/2)$, located between $(1,n)$ and
$(1,n+1)$, is chosen to be the origin for convenience. We search for
eigenstates with a Bloch wavevector $k$ along the zigzag edge and we
demand that the frequency should be equal to the single particle
multipole resonance frequency. If we truncate the Green function to
the nearest neighbor, the condition for the frequency to be equal to
the single particle multipole resonance frequency is "zero-hopping",
namely a zero sum of the fields from the three nearest neighbors.
This gives
\begin{equation}
\mathbf{G}(-\mathbf{r}_{1,n})\mathbf{V}_{1,n}+\mathbf{G}(-\mathbf{r}_{1,n+1})%
\mathbf{V}_{1,n+1}+\mathbf{G}(-\mathbf{r}_{2,n+\frac{1}{2}})\mathbf{V}_{2,n+%
\frac{1}{2}}=0,
\end{equation}%
where $\mathbf{V}_{m,n}=\mathbf{V}_{m}e^{inka}$ if the periodic Bloch
condition is imposed. From Eq. (1) a matrix $\mathbf{T}$ relating $\mathbf{V}%
_{1}$ and $\mathbf{V}_{2}$ can be defined as
\begin{equation}
\mathbf{T}=-\mathbf{G}^{-1}(-\mathbf{r}_{2,n+\frac{1}{2}})[\mathbf{G}(-%
\mathbf{r}_{1,n})e^{-ika/2}+\mathbf{G}(-\mathbf{r}_{1,n+1})e^{ika/2}].
\end{equation}%
This procedure can be done iteratively to obtain $\mathbf{V}_{m}=\mathbf{T}%
^{m-1}\mathbf{V}_{1}$. The eigenvalues of the matrix $\mathbf{T}$ give the
properties of the states. If $|\lambda |<1$, the states are localized edge
states, while $|\lambda |=1$ corresponds to a Bloch state extending to the
interior of the ribbon.

For out-of-plane modes, the Green function $\mathbf{G}^{P_{out}}(\mathbf{r}%
-\mathbf{r}^{\prime })$ is just an isotropic scalar function and Eq. (1) is
reduced to $p_{z1}e^{inka}+p_{z1}e^{i(n+1)ka}+p_{z2}e^{i(n+\frac{1}{2})ka}=0$%
, which is exactly the same condition found for electronic graphene \cite%
{nak:96}. This is why the edge state behavior of the out-of-plane
mode is so similar to that of electronic graphene ribbons. The ratio
$p_{z2}/p_{z1}$ is simply given by $-2\cos ka$ and is shown by green
lines in Fig. 4(a). For the in-plane dipole mode, the QSA dyadic
dipolar Green function can be
written as $\mathbf{G}^{p_{in}}(\mathbf{r}-\mathbf{r}^{\prime })\mathbf{P}%
_{in}(\mathbf{r}^{\prime })=\left[ 3(\mathbf{n}\cdot \mathbf{P}_{in})\mathbf{%
n}-\mathbf{P}_{in}\right] /\left\vert \mathbf{r}-\mathbf{r}^{\prime
}\right\vert ^{3}$, where $\mathbf{n}$ is the unit vector along the $\mathbf{%
r}-\mathbf{r}^{\prime }$ direction, and $\mathbf{T}$ is a $2\times
2$ matrix that can be found analytically. In Fig. 4(a), the absolute
values of the two eigenvalues are plotted by red and black lines.
The black curve is always larger than 1, while the red curve is
smaller than 1 when $ka<2\pi /3$, denoted as $\mathbf{P}_{1}$ in the
figure, which explains why in-plane dipolar edge states are confined
by $0\leq ka<2\pi /3$. For quadrupoles, there are five independent
variables $\{Q_{yy},Q_{zz},Q_{xy},Q_{xz},Q_{yz}\}$ and the
corresponding QSA Green function can be derived analytically after
tedious algebra. The details will be published elsewhere.
$\mathbf{T}$-matrix becomes a $5\times 5$ matrix, and the absolute
values of the five eigenvalues are plotted in Fig. 4(b). Only two
sets of eigenvalues give absolute values less than 1 and thus form
edge states. The edge state solution labeled as $\mathbf{Q}_{1}$
(blue) is smaller than unity in the whole 1D BZ. This corroborates
the existence of an edge state band that goes from $ka=0$ to $\pi $
in Fig. 2(c) in the $N=20$ zigzag plasmonic ribbon. Another state
denoted by $\mathbf{Q}_{2}$ (red) has a critical value
for the edge state at $ka=2\pi /3$ similar to the in-plane dipole mode $%
\mathbf{P}_{1}$.

\begin{figure}[tbp]
\centerline{\includegraphics[width=3in]{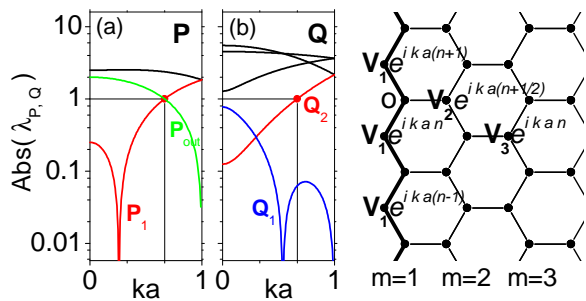}}
\caption{(Color online) Absolute values of eigenvalues $\protect\lambda $ of
the matrix $\mathbf{T}$ as a function of the Bloch wavevector $k$. $\mathbf{V%
}$ is either $\mathbf{P}$ or $\mathbf{Q}$. $|\protect\lambda |<1$ gives the
condition for edge states. (a) Dipole modes. The green line is for
out-of-plane modes and the red and black lines are for in-plane modes. (b)
Quadrupole modes. $\mathbf{Q}_{1}$ (blue) is always an edge state, while $%
\mathbf{P}_{1}$ and $\mathbf{Q}_{2}$ give edge state solutions when $ka<2%
\protect\pi /3$. The inset is a sketch of a semi-infinite honeycomb
structure with a zigzag edge. The indices $(m,n)$ and vector wave
functions are defined as shown.} \label{fig:fig4}
\end{figure}

Another interesting observation that can be deduced from Fig. 4 is that the
eigenvalues for the $\mathbf{T}$-matrix go to zero at $ka=0.26\pi $ and $%
0.54\pi $ for $\mathbf{P}_{1}$ and $\mathbf{Q}_{1}$ respectively, which
implies that the corresponding edge states are localized at the edge
completely. In contrast, completely localized states in electronic graphene
are only found at the BZ boundary $ka=\pi $. The completely localized state
away from the BZ boundary comes from the vector nature of EM waves.

In summary, we have found Dirac spectra in "graphene-like" plasmonic
structures for both dipole and quadrupole coupling. Edge states for
plasmonic lattices with finite widths are also identified and
analyzed numerically and their existence conditions are derived
analytically. The edge states for the out-of-plane dipole modes in
zigzag plasmonic ribbons resemble those of the electronic grapheme
zigzag edge states, but the edge states for in-plane dipole and
quadrupole modes have rather unique characters due to their vector
nature. These results can lead to the study of Dirac point behavior
in yet another class of interesting material.

This work was supported by the Central Allocation Grant from the Hong Kong
RGC through HKUST3/06C. Computation resources were supported by the Shun
Hing Education and Charity Fund. We thank Drs. Xianyu Ao and Kin Hung Fung
for helpful discussions.


\begin{thebibliography}{99}
\bibitem{kre:95} U. Kreibig and M. Vollmer, \textit{Optical Properties of
Metal Clusters} (Springer, New York, 1995).

\bibitem{mai:07} S. A. Maier, \textit{Plasmonics: Fundamentals and
Applications} (Springer, New York, 2007).

\bibitem{nie:97} S. Nie and S. R. Emory, Science \textbf{275}, 1102 (1997).

\bibitem{kne:97} K. Kneipp, \emph{et al.}, Phys. Rev. Lett. \textbf{78},
1667 (1997).

\bibitem{ang:06} P. Anger, P. Bharadwaj, and L. Novotny, Phys. Rev. Lett.
\textbf{96}, 113002 (2006).

\bibitem{mai:03} S. A. Maier, \emph{et al.}, Nat. Mater. \textbf{2}, 229
(2003).

\bibitem{tao:07} A. Tao, P. Sinsermsuksakul, and P. D. Yang, Nature Nano.
\textbf{2}, 435 (2007).

\bibitem{zhang:00} W. Y. Zhang, \emph{et al.}, Phys. Rev. Lett. \textbf{84},
2853 (2000).

\bibitem{nov:04} K. S. Novoselov, \emph{et al.}, Science \textbf{306}, 666
(2004).

\bibitem{loui:06} Y. W. Son, M. L. Cohen, and S. G. Louie, Nature (London)
\textbf{444}, 347 (2006).

\bibitem{nak:96} K. Nakada, M. Fujita, G. Dresselhaus, and M. S.
Dresselhaus, Phys. Rev. B \textbf{54}, 17954 (1996).

\bibitem{mara:91} M. Plihal, and A. A. Maradudin, Phys. Rev. B \textbf{44},
8565 (1991).

\bibitem{hal:08} F. D. M. Haldane and S. Raghu, Phys. Rev. Lett. \textbf{100}%
, 013904 (2008).

\bibitem{wang:08} Z. Wang, Y. D. Chong, J. D. Joannopoulos, and M. Soljacic,
Phys. Rev. Lett. \textbf{100}, 013905 (2008).

\bibitem{park:04} S. Y. Park and D. Stroud, Phys. Rev. B \textbf{69}, 125418
(2004); W. H. Weber and G. W. Ford, Phys. Rev. B \textbf{70}, 125429 (2004);
D. S. Citrin, Nano Lett. \textbf{4}, 1561 (2004); A. Alu and N. Engheta,
Phys. Rev. B \textbf{74}, 205436 (2006); A. F. Koenderink and A. Polman,
Phys. Rev. B \textbf{74}, 033402 (2006); V. A. Markel and A. K. Sarychev,
Phys. Rev. B \textbf{75}, 085426 (2007).

\bibitem{fung:07} K. H. Fung and C. T. Chan, Opt. Lett. \textbf{32}, 973
(2007); Y. R. Zhen, K. H. Fung, and C. T. Chan, Phys. Rev. B \textbf{78},
035419 (2008).
\end{thebibliography}
\end{document}